\documentclass[twocolumn,showpacs,preprintnumbers,amsmath,amssymb,superscriptaddress,prb]{revtex4-1}
\usepackage{epsfig,amsopn}
\usepackage{graphicx}
\usepackage{epstopdf}
\usepackage{sidecap}
\usepackage{color}
\usepackage{caption}
\usepackage{subcaption}  
\usepackage{amsmath,amssymb}
\usepackage{amsthm}
\usepackage{amsfonts}
\usepackage{textcomp}
\usepackage{enumerate}
\usepackage{soul}
\usepackage{float}
\usepackage{scalerel}
\usepackage{braket}
\usepackage{tikz}
\usepackage[titletoc]{appendix}
\usepackage[ colorlinks = true,
linkcolor = red,
urlcolor  = red,
citecolor = blue,
anchorcolor = red]{hyperref}
\newcommand{\beq}{\begin{equation}}
  \newcommand{\eeq}{\end{equation}}
\newcommand{\bqa}{\begin{eqnarray}}
  \newcommand{\eqa}{\end{eqnarray}}
\newcommand{\no}{\nonumber}

\newcommand{\kk} {\ensuremath{{\bf k}}}
\newcommand{\qq} {\ensuremath{{\bf q}}}

\newcommand{\QQ} {\ensuremath{{\bf Q}}}

\newcommand{\moire}{moir$\acute{e}\;$}
\newcommand{\tr}{\textrm}
\begin{document}
	\title{Excitonic Metal and Non-Fermi Liquid Behaviour in
          Twisted Double Bilayer Graphene near Charge Neutrality}
	\author{Unmesh Ghorai}
	\affiliation{Department of Theoretical Physics, Tata Institute of Fundamental Research,
          Homi Bhabha Road, Mumbai 400005, India }
        \author{Ayan Ghosh}
        \affiliation{ Physics Department, Indian Institute of Science,
          Bengaluru 560012, India}
        \author{Anindya Das}
        \affiliation{ Physics Department, Indian Institute of Science,
          Bengaluru 560012, India}
	\author{Rajdeep Sensarma}
	\affiliation{Department of Theoretical Physics, Tata Institute of Fundamental Research,
		Homi Bhabha Road, Mumbai 400005, India }
	%
	%\date{\today}
	%
	%
	\begin{abstract}
		Twisted double bilayer graphene is a compensated
                semi-metal near the charge neutrality point with the
                presence of small electron and hole pockets in its
                band structure. We show that strong Coulomb attraction
                between the electrons and holes can lead to the formation of
                indirect excitons. Condensation of these excitons at low temperature creates an
                excitonic metal with charge density wave order for
                an appropriate range of interaction strength. This has
                interesting implications for low-temperature transport
                in the system as a function of carrier density and
                temperature. The reorganization of the single particle
                excitations and their density of states in the
                excitonic metal can lead to peaks in
                resistivity as a function of carrier density, 
                recently seen in experiments at low temperatures. The
                fluctuations of the Landau damped order parameter in
                the quantum critical metal lead to non-Fermi liquid
                behaviour, which can explain the sublinear $T^{2/3}$
                dependence of the resistance near the
                charge neutrality point. 
	\end{abstract}

	\maketitle
	% \tableofcontents
	% \newpage
        
	\section{Introduction}

        When sheets of two-dimensional materials are stacked on top of
        each other, and their crystal axes are twisted
        (rotated) by a small angle, the electronic structure
        of these heterostructures become extremely sensitive to the angle of twist
        between them~\cite{BistritzerMacDonald2011,LopesdosSantos2012,KoshinoYuanKoretsuneOchiKurokiFu2018,LopesdosSantos2007, Bernevig2021,CarrMassattFangCazeauxLuskinKaxiras2017,Tarnopolsky2019,Zou2018,ChebroluChittariJung2019,Mohan2021}. This has led to the idea of twistronics~\cite{CarrMassattFangCazeauxLuskinKaxiras2017,CFDFTLSWTKAJ2018}, where
        the twist angle will be used as an experimental knob to change
        the electronic properties of these systems. The tunability of
        electronic structure with twist angles has been
        successfully studied in a controlled fashion in several
        systems,
        including multiple layers of graphene~\cite{CFDFTLSWTKAJ2018,CFFWTKJ2018,Yankowitz1059,Park2021,Hao2021,pablo_multi_2021} and Bernal stacked bilayer
        graphene~\cite{Cao_2020,BZTWMT2019,AdakSinhaGhoraiSanganiVarmaWatanabeTaniguchiSensarmaDeshmukh2020}, graphene-Boron Nitride 
        sandwiches~\cite{Mishchenko2014},
        and heterostructures
        made of dichalcogenides~\cite{Zhang2020,Wang}. A common feature
        of these
        systems is the presence of
        (multiple) magic twist angles~\cite{BistritzerMacDonald2011, Tarnopolsky2019}, where the bandwidth of the
        system is a minimum and the fate of the system is determined
        by strong electronic interactions. In graphene-based systems,
        such magic angles occur around
        $1-1.6^\circ$~\cite{Lu2019, LHKLRYNWTVK2020,Park2021},
        leading to large
        incommensurate \moire unit cells $\sim 8-15~ \textrm{nm}$.

        In twisted
        bilayer graphene (tBLG), where two sheets of graphene are twisted
        with respect to each other, the electronic interactions lead to a
        plethora of symmetry broken phases as a function of the
        electron density at the magic angle, from correlated insulators~\cite{CFDFTLSWTKAJ2018,Lu2019,Nuckolls2020} to orbital
        ferromagnets~\cite{Sharpe2019} to superconductors~\cite{CFFWTKJ2018,Yankowitz1059,Lu2019,Oh2021,Saito}. In twisted trilayer graphene (tTLG),
        where three sheets of graphene are twisted with respect to
        each other, the electronic correlations again lead to a
        superconducting state over a range of carrier densities~\cite{Park2021,Cao_ttlg_2021,Hao2021,Chen2019}. In
        comparison, their close cousin, the twisted double bilayer
        graphene (tDBLG), where two sheets of Bernal stacked bilayer
        graphene is rotated with respect to each other, had shown
        simple metallic behaviour as a function of filling in early
        experiments~\cite{AdakSinhaGhoraiSanganiVarmaWatanabeTaniguchiSensarmaDeshmukh2020,Cao_2020}. While there is evidence
        of correlated behaviour in
        presence of perpendicular electric~\cite{BZTWMT2019,LHKLRYNWTVK2020,Cao_2020,Burg2020} or magnetic fields~\cite{Cao_2020,Burg2020}, the
        phenomenology of plain vanilla tDBLG seemed to be explained by
        a non-interacting picture ~\cite{Cao_2020,AdakSinhaGhoraiSanganiVarmaWatanabeTaniguchiSensarmaDeshmukh2020,ChebroluChittariJung2019,Mohan2021}.

        The metallic behaviour of tDBLG even near the charge
        neutrality point (CNP) with no external doping is readily
        explained by the fact that the flat valence and conduction
        bands in tDBLG overlap with each other in energy. This leads to
        the formation of small electron and hole pockets in this
        regime~\cite{ChebroluChittariJung2019, Mohan2021,koshino2019}. tDBLG at CNP is thus a compensated
        semi-metal. The presence of these  electron and hole pockets near CNP in
        tDBLG has now been
        demonstrated unambiguously through recent magnetotransport
        measurements~\cite{anindya_thermo}. This brings us to an
        interesting question: Do the strong electronic interactions
        have any qualitative effect on the small electron and hole
        pockets near CNP, or do they behave like almost
        non-interacting systems? In
        this paper, we show that indirect excitons are formed due to the attraction between the electron
        and hole pockets, and
       condensation of these excitons can lead to the formation of a CDW state near CNP
        at low temperatures. The
        system remains metallic on either side of the transition for a range of parameters. The
        phase transition in the background of itinerant electrons is
        driven by Landau damped fluctuations of the excitonic
        order. The scattering of electrons by these fluctuations lead
        to non-Fermi liquid behaviour in these systems~\cite{Metlitski2010, Hertz_qcp,millis_qcp,subir_antifer_qcp,Vekhter2004,Senthil2004,Polchinski,Senthil_vojta_2004,Metlitski2}. Thus
        interactions have profound effects on the small electron and
        hole pockets in the system.

        The
        formation of excitonic condensate leads to a reorganization
        of the electronic structure into multiple ``minibands'' with
        their respective Fermi seas. This leads to anomalous peaks in
        the inverse density of states at Fermi level as a function of
        carrier density, which mimics the peaks in resistance as a
        function of carrier density seen in
        recent experiments~\cite{anindya_thermo}. Within a mean field
        theory,
        the simultaneous
        presence of the anomalous peaks as well as a Fermi sea to
        account for metallic behaviour strongly constrains the
        interaction parameters. We find that the allowed parameter ranges
        are reasonable for tDBLG.

       Another surprising result from these experiments~\cite{anindya_thermo} is that close
       to CNP, the measured resistance 
        exhibits a unique sublinear temperature dependence ($R\sim
        T^{2/3}$) in the temperature range
        $0-10~\textrm{K}$. Above this temperature range, the
        resistance increases linearly with temperature. Far away
        from the CNP, the resistance reverts to
        a standard superlinear temperature dependence $(R\sim T^2)$
        seen in usual metals. The sublinear temperature dependence of
        resistivity is rarely seen in metals and  cannot be explained by
        standard scattering mechanisms (disorder, electron-electron or
        electron-phonon)
        within a Fermi liquid
        theory. On the other hand, in a quantum critical metal on the verge of forming
        excitonic condensates, the quantum fluctuations of the order
        parameter will be Landau damped by the low energy
        Fermions present in the system. We show that the scattering of
        charge carriers
        in the small Fermi
        pockets from such overdamped critical
        fluctuations gives rise to a non-Fermi liquid with a $T^{2/3}$ dependence of resistance at
        low temperatures. At higher temperatures, we reach an
        equivalent of a ``Bloch Gruneissen'' temperature for these
        fluctuations and the resistivity shows linear temperature
        dependence beyond that scale. The question of a breakdown of Fermi liquid
        theory in a metal due to quantum fluctuations near a critical
        point is a matter of great theoretical interest, and has been
        studied using sophisticated formalisms~\cite{Metlitski2010,Metlitski2,subir_antifer_qcp,Senthil2004,Vekhter2004,Senthil_vojta_2004,Polchinski}. Here we propose that
        the presence of small Fermi surfaces in these systems
        undergoing phase transitions make sure that the non-Fermi
        liquid behaviour shows up in low temperature transport as a
        non-analytic temperature dependence of the resistivity. Thus
        we propose that the magic angle tDBLG is
       not a garden-variety metal; rather there is strong
       experimental evidence for an underlying non-Fermi liquid state
       formed due to interactions in a compensated semi-metal.

       In this paper, we review the formation of electron and hole
       pockets in tDBLG in Sec.~\ref{ehpocket}. Sec.~\ref{excmft}
       provides the details of the mean field theory of the excitonic
       condensate near CNP. In Sec.~\ref{respeak}, we show how the
       electronic reorganization due to exciton formation leads to
       peaks in the inverse density of states and compare it to
       experiments. In Sec.~\ref{nfl}, we discuss how the Landau
       damping near a critical point leads to a non-Fermi liquid
       behaviour. We also discuss the connection of this underlying
       non-Fermi liquid state to the sublinear temperature dependence
       in resistivity.  Finally in Sec.~\ref{concl}, we conclude with
       a summary of our results.

      \section{Electron and Hole pockets in tDBLG \label{ehpocket}}

        The low energy electronic states of tDBLG at magic angle at
        and around the CNP consist of both electron and
        hole pockets. This is in contrast with other members of
        the twisted graphene family (tBLG or tTLG) which forms a Dirac
        node at CNP and shows either electron or hole like states
        when doped away from the CNP~\cite{BistritzerMacDonald2011,KoshinoYuanKoretsuneOchiKurokiFu2018,Subir_ttlg_2022,macdonal_trilayer_2019,koshino2019,Mohan2021}. As a result, tDBLG shows metallic
        behaviour near CNP, while the other \moire graphene systems
        show weakly insulating ($R\sim10-50 \,\tr{k}\Omega$) behaviour~\cite{CFDFTLSWTKAJ2018,LHKLRYNWTVK2020,SCWLWZTLTWTYMSYZ2020,AdakSinhaGhoraiSanganiVarmaWatanabeTaniguchiSensarmaDeshmukh2020,Cao_ttlg_2021,Park2021,Yankowitz1059,Hao2021}.

\begin{figure}
        	\includegraphics[width=\columnwidth]{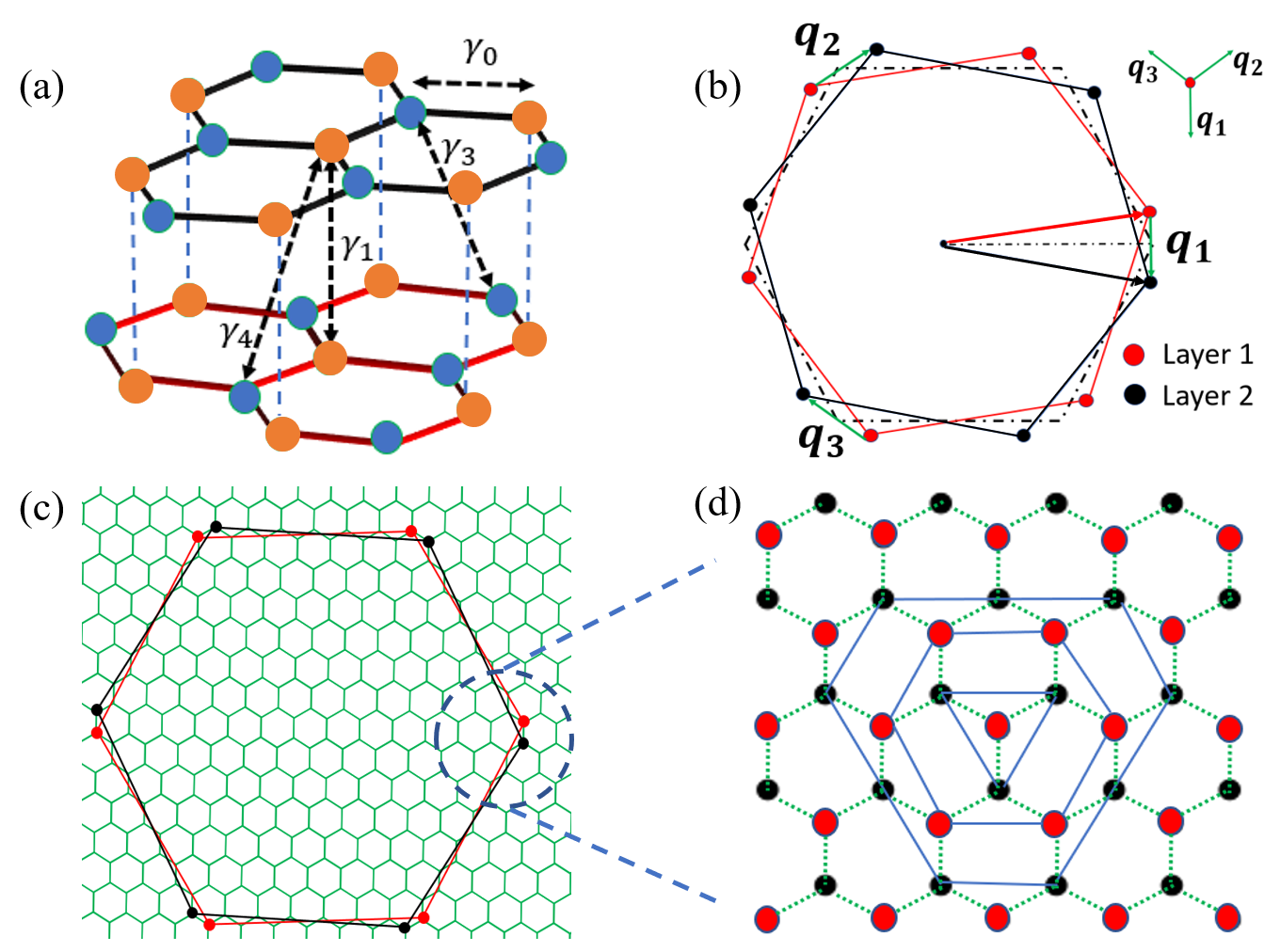}
        	\caption{(a) Schematic of Bernal stacked bilayer
                  graphene and relevant intra ($\gamma_0$) and
                  inter($\gamma_{1(3)(4)}$)-layer tunnelings. (b)
                  Brillouin zone of two bilayer-graphene sheets
                  rotated w.r.t each other by a small twist
                  angle. Three transfer momenta ($q_{1(2)(3)}$)
                  between the Dirac points
                  of each layer are marked. (c) and (d) show the
                  smaller \moire
                  Brillouin zone created by the twist. The zoomed-in
                  image
                  shows the construction of consecutive shells and the
                  momentum
                  cut-off used in the continuum model.}
        	\label{Fig:Fig_1}
        \end{figure}

        Since tDBLG is made of two layers of bilayer graphene (BLG)
        with a twist between them, it is useful to start with the band
        structure of BLG. A schematic of a Bernal (A-B) stacked BLG is
        shown in Fig.~\ref{Fig:Fig_1}(a). The carbon atoms in each layer
        form a honeycomb lattice and are
        coupled by a nearest neighbour in-plane hopping $\gamma_0$ where, $ \frac{\sqrt{3}}{2}\gamma_0=\hbar v_0/a \sim
        2.1354 ~ \tr{eV}$~\cite{koshino2019}, while the out of plane hopping is primarily between
        atoms which sit on top of each other (in
        the so-called dimer sites)  with a scale $\gamma_1 =400
        ~\tr{meV}$. A simplified model of BLG with only $\gamma_0$ and
        $\gamma_1$ produces a Dirac point where two quadratic bands
        touch each other, with a Berry monopole of charge $2$ located
        at the Dirac point. However, if one considers additional
        interlayer hoppings between the non-dimerized sites (the
        trigonal warping $\gamma_3= 320~\tr{meV}$), or between dimerized and
        non-dimerized sites ($\gamma_4 = 44~\tr{meV}$), this simple picture
        changes at low energies. The single band touching point splits into a
        central Dirac point and three satellite Dirac points, where
        linearly dispersing bands touch each other~\cite{McCann_2013}.  The central and the
        satellite Dirac points carry Berry monopoles of opposite
        charges. The formation of the satellite Dirac points and
        associated Lifshitz transitions in BLG have been probed both
        theoretically and experimentally~\cite{Falko_liftshitz_2010,varlet_liftshitz}. We also include an on-site potential of dimerized sites with respect to non-dimerized sites, $\delta'=50\,\text{meV}$~\cite{koshino2019}.

        We consider the $AB-AB$ stacked tDBLG.
        In this case, the graphene Brillouin zone is tiled by
        the \moire Brillouin zone (mBZ), with a reciprocal lattice vector of
        size $k_M=k_D \sin \theta/2 $, where $k_D$ is
          the reciprocal lattice vector of the BLG Brillouin zone and
          $\theta$ is the twist angle, as shown in Fig.~\ref{Fig:Fig_1}(b).
          The
          two twisted layers are tunnel-coupled with tunneling between
          the same sublattices $u_{AA}=u_{BB}=79.7 ~ \tr{meV}$ and the
          tunneling between different sublattices $u_{AB}= 97.5 ~
          \tr{meV}$. For these set of parameters, used in a wide range
          of earlier papers~\cite{koshino2019,LHKLRYNWTVK2020,Mohan2021}, the magic angle is
          $1.2^\circ $, which matches with experimental estimates of
          magic angle for tDBLG~\cite{anindya_thermo}. The inter-BLG tunnelings couple momentum states
          in one mBZ to those in nearby zones; as shown in
          Ref.~\onlinecite{BistritzerMacDonald2011} and Fig.~\ref{Fig:Fig_1}(c-d). One can limit the number of
          Brillouin zones used to calculate the band dispersion of the
          \moire system at low twist angles~\cite{BistritzerMacDonald2011}. We use the 5 nearest shells which lead to a $184$
          dimesional continuum Hamiltonian to obtain the low energy
          band dispersion of tDBLG within an accuracy of $1\%$.

\begin{figure*}[ht]
	\includegraphics[width=\textwidth]{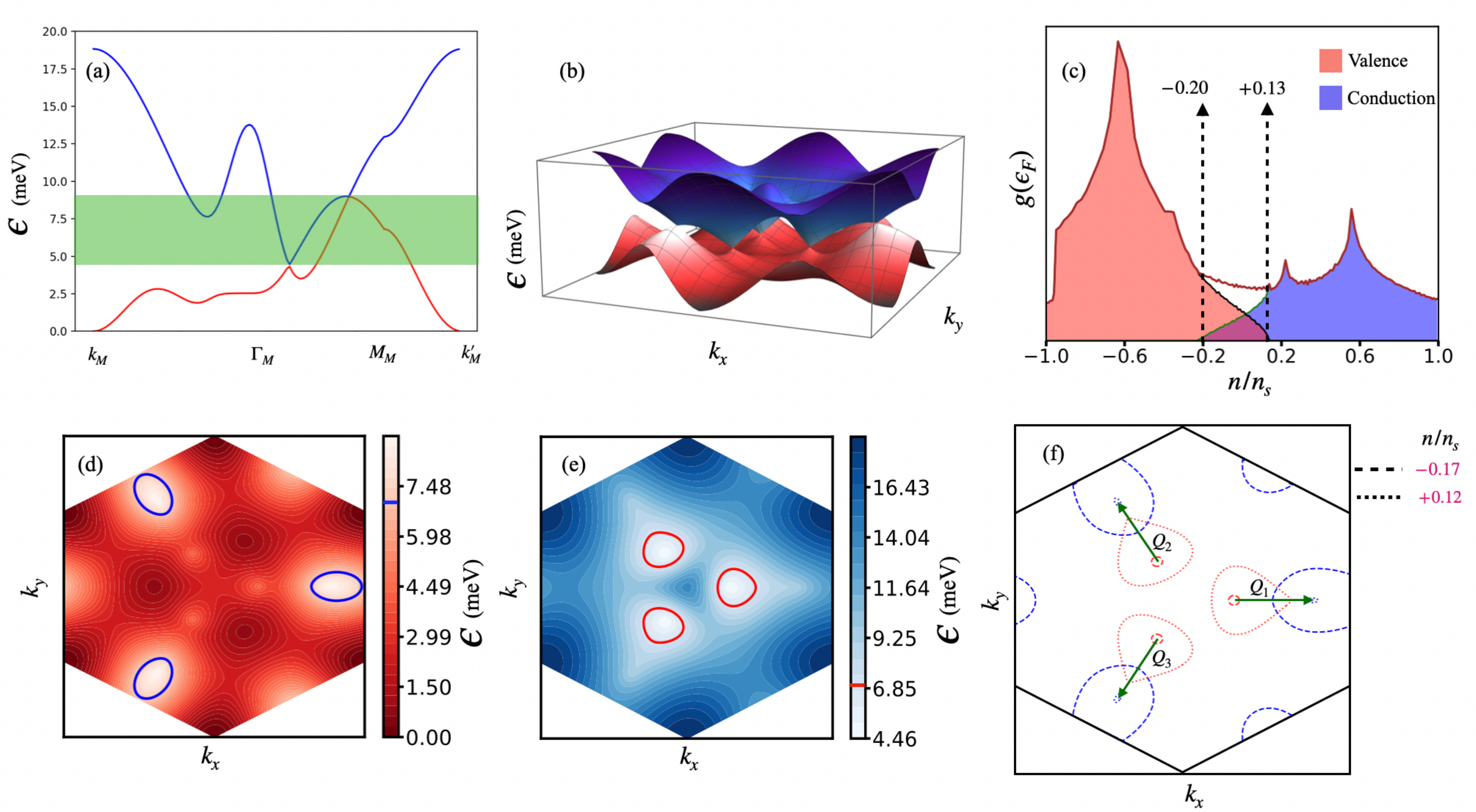}
	\caption{(a) Non-interacting band dispersion of tDBLG along
          high symmetry directions in the moire Brillouin zone. The
          valence (red) and conduction (blue) bands are plotted. The
          shaded region denotes the energies where the bands
          overlap. (b) A 3-d depiction of the energy dispersion in the
          mBZ to show the six satellite Dirac points around the
          $\Gamma_M$ point. (c) Valence (red) and conduction (blue)
          bands' density of states at the Fermi level are plotted with
          the normalised carrier density. The bands overlap between
          carrier densities (marked)$\approx$ -0.20 and +0.13. (d) and
          (e) Contour plots of the valence (red) and conduction(blue)
          bands respectively with Fermi surfaces at CNP marked
          separately. The blue pocket in (d) and the red pocket in (e)
          denote the hole and electron pocket respectively. (f) The
          electron (red) and hole (blue) pockets at two different
          carrier densities around the CNP are shown to point out that
          one of the pockets shrinks and another one inflates as one
          moves away from CNP. The $Q_{1(2)(3)}$ vectors connect the
          e-h pocket
          centers and are related by $C_3$ symmetry.}
	\label{Fig:Fig_2}
\end{figure*}

          The low energy band structure of tDBLG near the magic angle consists of
          a valence and a conduction band with a bandwidth $\sim 20\,
          \tr{meV}$.  In Fig.~\ref{Fig:Fig_2}(a), we plot the dispersion of the
          conduction and valence bands along the principal directions of the
         mBZ ($K_M-\Gamma_M-M_M-K'_M$). We clearly see that the
          bands overlap in energy along the $\Gamma_M-M_M$ line (shown by
          the shaded region in Fig.~\ref{Fig:Fig_2}(a)) ; hence at the charge neutrality
          point, one would expect a compensated metal with electron
          and hole pockets. We would like to note that particle-hole
          symmetric band structures, obtained from a simplified description of
          BLG with only $\gamma_0$ and $\gamma_1$, do not show this
          band overlap~\cite{Mohan2021,ChebroluChittariJung2019}. The presence of trigonal warping $\gamma_3$ is
          crucial in obtaining this overlap. In Fig.~\ref{Fig:Fig_2}(b), we show a 3d plot
          of the dispersion of the valence and the conduction band, which
          clearly shows that the bands do not cross each other; rather
          they touch each other at two anisotropic Dirac points along
          the $\Gamma-M$ line at slightly different energies, leading
          to the band overlap. The valence and conduction band
          dispersions are plotted as color plots in the full Brillouin
          zone in Fig~\ref{Fig:Fig_2}(d) and (e). The thick lines mark the
          Fermi surfaces (curves) of the electron (conduction band)
          and hole (valence band) pockets at the CNP. We clearly see that there are three electron and
          three hole pockets related by $C_3$ symmetry. The pockets
          are centered around the points where the bands touch each
          other. We note that we have plotted the moire bands around
          one valley of the original BLG dispersion. The band
          touchings and electron-hole pockets of the other valley can
          be obtained by applying a rotation of $\pi$ to this figure.

          As we move away from the CNP on the electron doped side, the
          size of the electron pocket increases, while the hole pocket
          shrinks. At an electron density $n/n_s\sim0.13$, the hole
          pocket shrinks to a point and beyond this density, the
          system only has three electron pockets. Here $n_s\approx 3.3 \times
          10^{12}cm^{-2}$ is the density where the conduction band is
          fully filled (including spin and valley degeneracies). On the other hand,
          with hole doping, the hole pocket grows and the electron
          pocket shrinks, till it disappears at
          $n/n_s\sim-0.2$. Fig~\ref{Fig:Fig_2}(f) plots the Fermi surfaces
          for the electron and hole pockets on either side of CNP at
          densities $n/n_s=-0.17$ and $n/n_s= 0.12$ to show the evolution
          described above. Note that the centers of the pockets do not
          change with density and the wave-vectors joining the centers of the nearest electron and hole
          pockets, $\QQ_1$, $\QQ_2$, and $\QQ_3$ are clearly shown in
          Fig~\ref{Fig:Fig_2}(f). While there are interesting evolution of
          the Fermi surface at higher densities, in this paper, we
          will focus on densities between $n/n_s=-0.2$ and
          $n/n_s=0.13$, where both electron and hole pockets are
          present.

          The finite density of states at the Fermi level coming from
          these electron and hole pockets lead to metallic behaviour
          in tDBLG~\cite{Cao_2020,AdakSinhaGhoraiSanganiVarmaWatanabeTaniguchiSensarmaDeshmukh2020,BZTWMT2019,LHKLRYNWTVK2020,SCWLWZTLTWTYMSYZ2020} in absence of perpendicular electric field. We note that the waxing and waning of the
          electron and hole pockets compensate each other to keep the total density of
          states at the Fermi level finite and independent of the
          carrier density near CNP, as shown in Fig~\ref{Fig:Fig_2}(c). Recently the
          strong magnetic field dependence of the low temperature
          resistance and thermopower in these systems~\cite{anindya_thermo} have provided
          concrete evidence of the existence of these electron and hole
          pockets in tDBLG.

          \section{Exciton Condensates near CNP\label{excmft}}

        An important question in systems with low electronic density
        is: what is the fate of the system when the strong electronic
        interactions are taken into account? While earlier experiments
        in tDBLG~\cite{Cao_2020,AdakSinhaGhoraiSanganiVarmaWatanabeTaniguchiSensarmaDeshmukh2020,BZTWMT2019,LHKLRYNWTVK2020,SCWLWZTLTWTYMSYZ2020} showed a fairly standard metallic behaviour in
        absence of perpendicular electric fields (correlated states
        were found at finite electric fields), a recent
        experiment~\cite{anindya_thermo} at very low temperatures ($T< 2 K$) has shown a
        double peak structure in the resistance as a function of
        carrier density near CNP where both electron and hole pockets
        are present. These double peak structures cannot be
        explained by a non-interacting theory, since electrons and
        holes contribute additively to the electrical response, and
        the total density of states near CNP is almost independent of
        doping as seen in Fig~\ref{Fig:Fig_2}(c). These are the first
        concrete experimental signatures
        that electronic correlations play an important role in tDBLG
        near the CNP.

        In systems with electron
        and hole states at the Fermi level, the Coulomb attraction
        between the oppositely charged electrons and holes often leads
        to the formation of charge neutral electron-hole pairs called
        excitons. Formation of exciton states are commonly seen in
        semiconductor systems~\cite{HALPERIN1968115,KHVESHCHENKO2004323}, as well as Van der Waals
        heterostructures~\cite{zhang_exc_moire,chen_exc_ins_moire}. Coherent condensation of these particle-hole
        pairs,  leading to superfluidity of charge neutral objects, have been predicted and demonstrated in
        bilayer quantum Hall systems~\cite{eisenstein2014,Wang_exc_cond_2019,anshul_exc_cond,Li_exc_SF_2017} as well as
        bilayer graphene~\cite{Ju2017}. The electron-hole pockets in
        tDBLG, which are separated by a small momentum $|\QQ_i| \sim
        0.01~ \tr{\AA}^{-1}$, are ideal candidates for the formation of
        excitonic condensates with finite momentum (indirect excitons). We will now explore this possibility within mean field theory.
         \begin{figure}
          \includegraphics[width=\columnwidth]{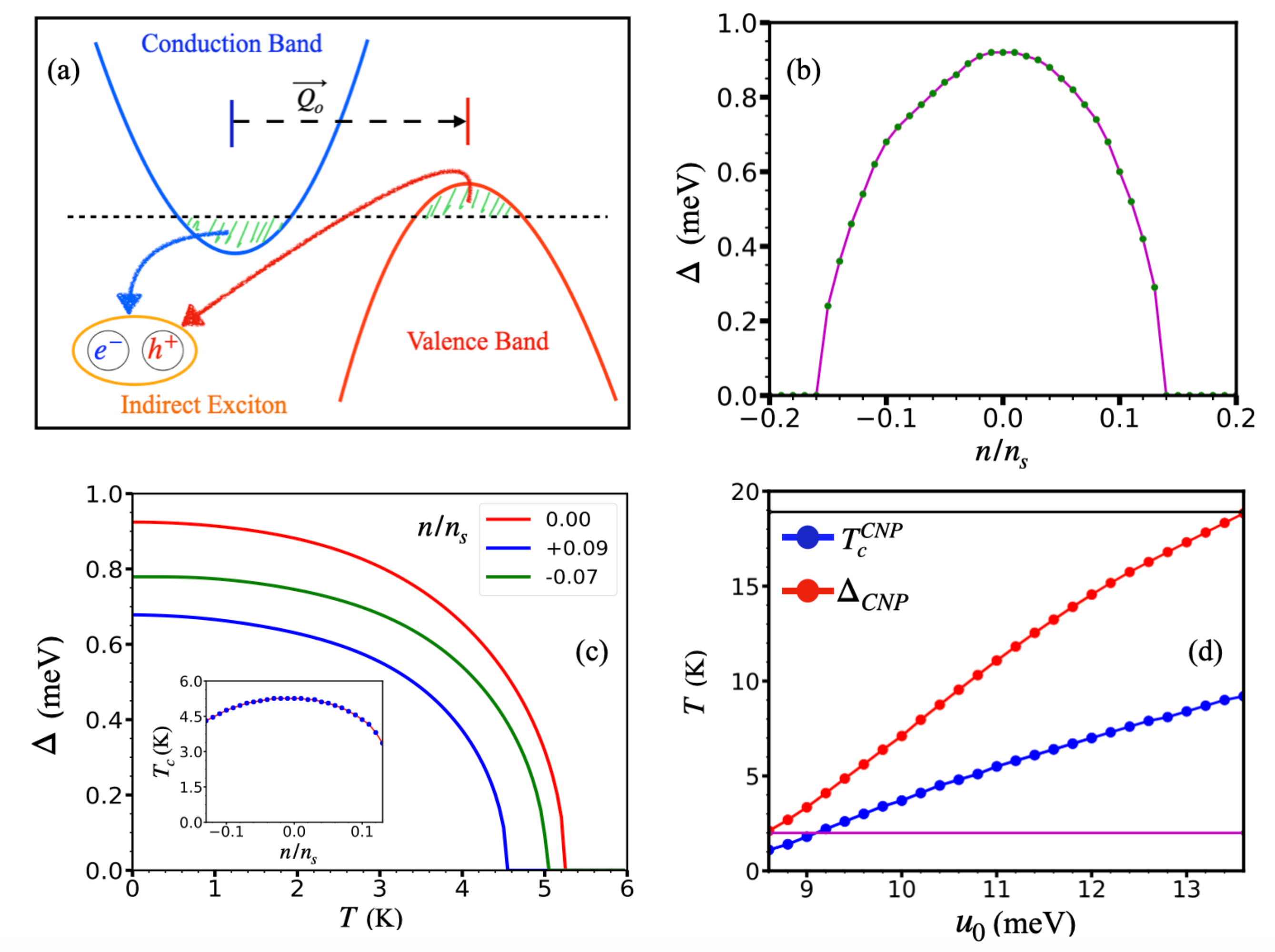}
          \caption{(a) Schematic showing the formation of indirect
            excitons from the two bands. (b) Mean field order
            parameter($\Delta$) is plotted with carrier
            density. $\Delta$ has a peak at CNP and goes down on
            either side of CNP. (c) The temperature dependence of
            $\Delta$ is shown for three values of $n/n_s=0.00,~0.09,
            ~-0.07$. In the inset, we have plotted the variation of
            $T_c$ with density around CNP. $T_c$ is weakly dependent
            on $n/n_s$ with a peak at CNP. (d) $\Delta$ (in the units
            of temperature) (red) and mean field $T_c$(blue) are plotted as a
            function of the interaction strength $u_0$. The pink line
            is the experimentally  observed $T_c$. The black line
            indicates $\Delta_c$, where the system forms an excitonic
            insulator rather than a metal for $\Delta >\Delta_c$. }
          \label{Fig:Fig_3}
        \end{figure}

        We consider the two-band Hamiltonian
	\begin{eqnarray}
	\no H&=&\sum_{\kk \eta}(\epsilon^c_{\kk}-\mu)C_{\kk,\eta}^{c\dagger}C_{\kk,\eta}^{c}+(\epsilon^v_{\kk}-\mu)C_{\kk,\eta}^{v\dagger}C_{\kk,\eta}^{v}\\
        &+&\frac{1}{\Omega}\sum_{\kk,\kk',\qq,\eta\eta'}V(\qq)C_{\kk+\qq,\eta}^{v\dagger}C_{\kk,\eta}^{v}C_{\kk'-\qq,\eta'}^{c\dagger}C_{\kk',\eta'}^{c} 
      \end{eqnarray}
      where $C^{c(v)}$ denotes the electron annihilation operator in
      the conduction (valence) band, $\eta$ is a composite spin and
      valley index and $\epsilon^{c(v)}_\kk$ is the
      conduction and valence band dispersions shown in the previous
      section. Here $V(\qq)$ is the screened Coulomb interaction between the conduction
      and valence band electrons. We note that we have neglected
      interaction between electrons in the same band, since our
      primary focus is on understanding the formation of interband excitons. We have also assumed that the large momentum
      connecting the two valleys of BLG $\sim 1 \tr{\AA}^{-1}$ makes it
      unfavourable for Coulomb interaction to scatter electrons
      from one valley to the other. Note that although excitons are
      often understood as pairing of electrons and holes, where the hole can be obtained by a particle-hole transformation on
      the valence band, i.e. $C^v \rightarrow h^{v\dagger}$, we prefer to
      work with electron coordinates in both bands. 
    The size of the small Fermi pockets as well as the momentum separating the
      center of the electron and hole pockets are $\sim 0.01
      \tr{\AA}^{-1}$. We assume that $V(\qq)$ does not change rapidly
      over this scale and can be approximated by a constant value
      $V_0$ for our calculation. For our calculations, we will use
      $u_0=10.9~\textrm{meV}$ where $u_0=V_0/\Omega$.  We will later try to constrain its value
      both from theoretical and experimental estimates.

      From
      Fig.~\ref{Fig:Fig_2}(f), we see that there are three electron and hole
      pockets in the mBZ, and three wave-vectors
      $\QQ_{1(2)(3)}$ connecting them. We will consider a mean-field
      description of finite momentum interband excitonic condensate,
      which has an equal amplitude at each of the momenta $\QQ_i$,
      i.e. $\frac{V_0}{\Omega}\sum_\kk \langle C^{\dagger c}_{\kk,\eta}
      C^v_{\kk+\QQ_i,\eta}\rangle = \Delta$ (See
      Fig.~\ref{Fig:Fig_3}(a) for a schematic of finite momentum exciton). We
      note that formation of finite momentum excitonic condensates in
      this system is equivalent to the appearance of a charge density
      wave (CDW) order in the system. Since $\sum_i\QQ_i =0$, our
      ansatz of equal condensates for all momenta ensure that there is
      no underlying valley current in the ground state of the
      system. It also ensures that the $C_3$ symmetry of each valley
      remains unbroken in the system.  Note that $\QQ_i \rightarrow -\QQ_i$, if the valley is
      flipped, and the net current summed over valleys would have been zero even for an 
      ansatz with unequal condensates. With this ansatz, using the
      basis $(C_\kk^c\;
      ,\;C_{\kk+\QQ_1}^v,\;C_{\kk'}^c,\;C_{\kk'+\QQ_2}^v,\;C_{\kk''}^c,\;C_{\kk''+\QQ_3}^v)$,
      we obtain the $6\times 6$ mean field Hamiltonian,
	\begin{equation}\label{meanfham6b6}
	\mathcal{H}=\begin{bmatrix}
	d_{\QQ_1}(\kk)&0&0\\ 0&d_{\QQ_2}(\kk')&0\\0&0&d_{\QQ_3}(\kk'')
	\end{bmatrix}
	\end{equation}
	where 
	\begin{equation}\label{meanfham_mod}
	d_{\QQ_i}(k)=\begin{bmatrix}
	\frac{\epsilon^c_{\kk}}{3}-\mu&\Delta\\ \Delta&\frac{\epsilon^v_{\kk+\QQ_i}}{3}-\mu
	\end{bmatrix}.
	\end{equation}
	Here $\kk'$ and $\kk''$ vectors are generated by rotating the $\kk$
        vector by $2\pi/3$ and $4\pi/3$  respectively. The
        quasiparticle excitation spectrum of the mean field theory is
        \begin{equation}
	E^{\pm}_{\QQ_i}(\kk)=\frac{\epsilon_\kk^c+\epsilon_{\kk+\QQ_i}^v}{6}-\mu \pm \sqrt{\frac{(\epsilon_\kk^c-\epsilon_{\kk+\QQ_i}^v)^2}{36}+\Delta^2}
	\label{disp_exc}
	\end{equation}
        The self-consistency equations, which determine the order
        parameter $\Delta$ and the chemical potential $\mu$, are given
        by
		\begin{eqnarray}\label{mfeq1}
		\no	1&=&\frac{V_0}{2\,\Omega}\sum_{\kk,\QQ_i}\frac{-1}{E_\kk^{\QQ_i}}[f(E_{\QQ_i}^{+}(\kk))-f(E_{\QQ_i}^{-}(\kk))];\\~n&=&\frac{1}{3}\sum_{\kk,\QQ_i}[f(E_{\QQ_i}^{+}(\kk))+f(E_{\QQ_i}^{-}(\kk)]\label{mfeq2}
		\end{eqnarray}
       where $f$ is the Fermi function. In Fig.~\ref{Fig:Fig_3}(b) we plot the self-consistent $\Delta$
       at $T=0$
       as a function of the carrier density $n/n_s$ for a system with
       $u_0=10.9~\textrm{meV}$. The order parameter shows a maxima
       around the CNP ($\sim 0.9~\textrm{meV}$) and decreases on
       either side of CNP, finally vanishing through a sharp jump at
       the boundaries where the electron-hole pockets cease to exist
       simultaneously. The electron and hole pockets are matched in
       size (albeit shifted in momentum) at the CNP. As we move away,
       the electron(hole) pocket grows while the hole (electron)
       pocket shrinks. This mismatch of the Fermi pockets leads to a
       weakening of the order parameter at finite carrier
       densities. The temperature dependence of the order parameter at
       three different densities are shown in
       Fig.~\ref{Fig:Fig_3}(c). From the vanishing of the
       order parameter, one can obtain the mean-field $T_c$ of the
       system, which is plotted in Fig.~\ref{Fig:Fig_3}(c) inset as a
       function of carrier density. We see that $T_c$ is very weakly
       dependent on carrier density and hovers around $5~\tr{K}$ for our
       chosen set of parameters. Our
       estimate is in the same ballpark as the experimentally obtained
       $T_C \sim 2~\tr{K}$~\cite{anindya_thermo}, where the resistance peaks disappear. We would like to note that this
       mean-field estimate is an upper bound for the real $T_c$, which
       will be further degraded by fluctuations and disorder.
		In Fig.~\ref{Fig:Fig_3}(d), we have shown the $u_0$ dependence of $\Delta(T=0)$ and $T_c$ at CNP. The magenta horizontal line is the experimentally observed $T_c\sim2\;\tr{K}$~\cite{anindya_thermo}. The black horizontal line corresponds to the maximum value of $\Delta(T=0)$ for which the system remains metallic (see discussion in Sec.~\ref{respeak} for details). These two lines thus provide lower and upper bounds on $u_0$ ($9.1\;\tr{meV}$ and $13.6\;\tr{meV}$ respectively) to be used in the calculation. We have used $u_0=10.9\;\tr{meV}$ in the middle of this range.
	
	 % Finally with all the self-consistent parameters, in FIG.~\ref{fermiS_Otemp} we plot the Fermi surface of the fermionic excitations mentioned in eq~\ref{disp_exc}. The finite Fermi surface near the CNP leads to metallic correlated states which can be verified through transport measurements. In the next section we have calculated the inverse density of states of these excitations and pointed out the signatures of it in the experimental data. Although the previous plots were made using zero T formulas, but in FIG.~\ref{fermiS_Otemp} we have plotted temperature dependence of the order parameter to focus on the fact that with increasing temperature the particles get more and more thermal energy and the exciton pair breaks down into electron and hole, thereby reducing the exciton order parameter to zero at some critical T ($T_c\sim5K$). \\ 

       \begin{figure*}[t]
         \includegraphics[width=\textwidth]{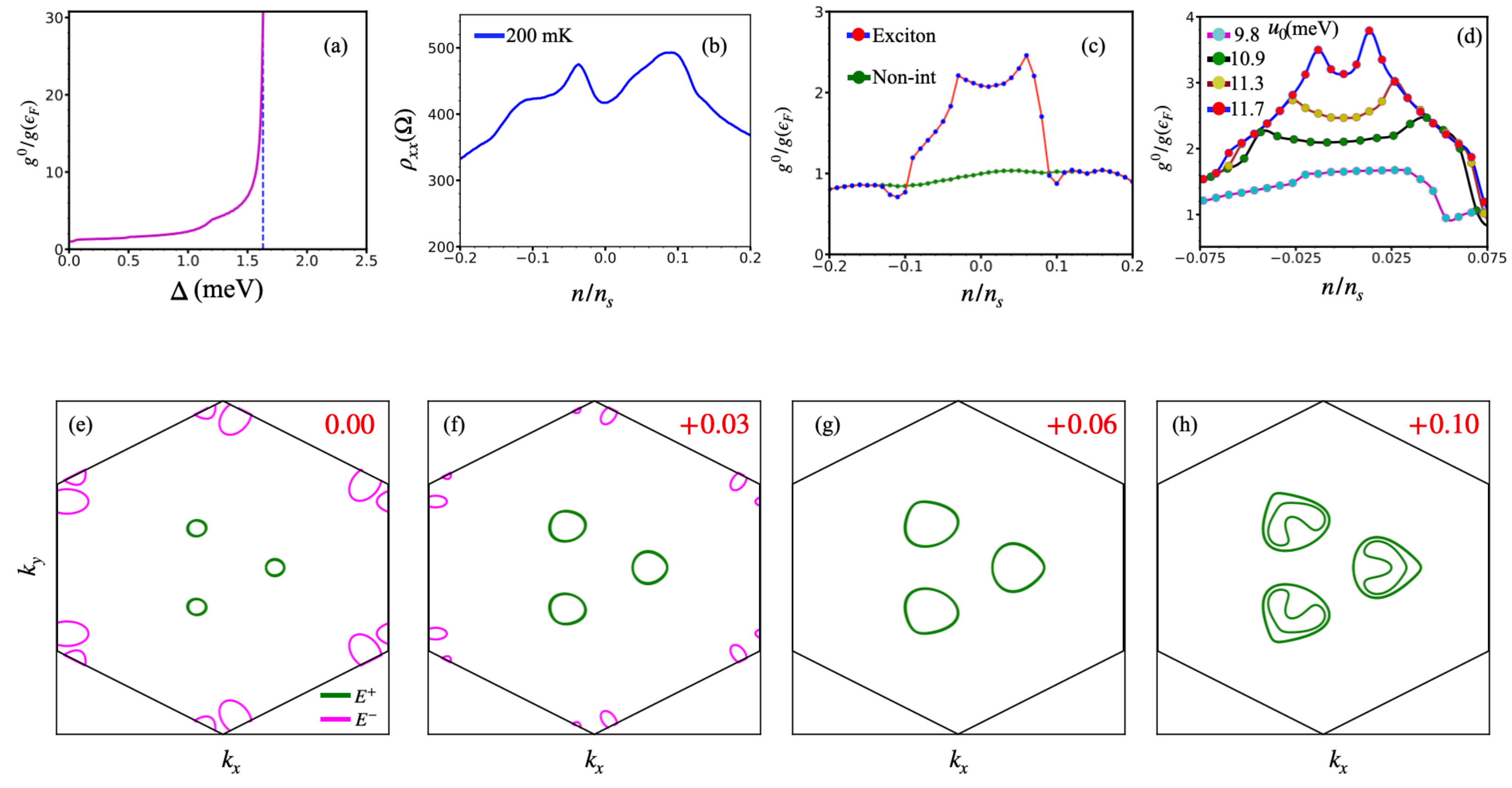}
         \caption{(a) Inverse DOS at the Fermi energy for the
           excitonic metal is plotted with
           $\Delta$. For $\Delta >\Delta_c \sim 1.6 ~\textrm{meV}$
           (marked by the blue dotted line),
           the system shows insulating behaviour with zero density
           of states at the Fermi level. (b)
           Experimentally obtained longitudinal resistivity of tDBLG as a
           function of carrier density at 200 mK exhibiting the double
           peak structure. (c) Inverse DOS at
           fermi level for the excitonic metal (red) and the
           non-interacting state (brown) as a function of carrier
           density. The excitonic condensate leads to two peaks in the
           inverse DOS similar to the experimental data in (b),
           whereas the non-interacting DOS at Fermi level is almost
           independent of density in this regime. (d) Inverse DOS at
           the Fermi level of the excitonic metal for different
           interaction strengths are plotted as a function of carrier
           density. The fact that the two peak structure disappears at
           low interaction strength, while the system is insulating at large
           interaction strength allows a small range of the
           interaction parameter which is compatible with the existing
           experimental data. We will use $u_0=10.9$ meV  for our
           calculations unless otherwise stated. (e-h) shows the
           evolution of the Fermi surfaces of the two exciton bands
           ($E^{+(-)}$) (green(pink)) in the mBZ with changing
           densities. For $0< n/n_s < +0.06$ the green contours
           increase while the pink contours shrink. This leads to an
           increase in the resistivity. At $n/n_s=0.10$,
           the Fermi level enters a
           higher miniband of the excitonic state. The additional
           density of states leads to a decrease in the resistivity,
           thus giving rise to the peak as a function of density. A
           similar story plays out on the other side of CNP, leading
           to the double-peak structure in resistivity.}
         \label{Fig4}
       \end{figure*}

	\section{Exciton condensates and Electronic Structure\label{respeak} }

        The occurrence of excitonic condensates in
        materials~\cite{KHVESHCHENKO2004323,Jia_2021,HALPERIN1968115,zhang_exc_moire} often lead to the underlying Fermi
        surface being gapped out, leading to an excitonic
        insulator. This is inevitably the case when the condensate is
        formed with $0$ net momentum, as is the case in quantum Hall
        bilayers and bilayer graphene~\cite{eisenstein2014,Wang_exc_cond_2019,anshul_exc_cond,Li_exc_SF_2017,Ju2017}. However, the shift between the
        center of the electron and hole pockets in tDBLG near CNP
        naturally leads to excitonic condensates with finite
        momenta. This leads to the possibility of having either
        metallic or insulating behaviour depending on the magnitude of
        the excitonic order in the system and the resulting
        quasiparticle spectrum $E^{\pm}_{\QQ_i}$. Note that the
        original 2-band system is mapped to a $6$ band system in the
        presence of excitonic order.

        A simple measure of
        metallicity of the system is the density of states at the
        Fermi level, $g(\epsilon_F)= \sum_{\pm,i,\kk}\delta\left(
          E^\pm_{\QQ_i}(\kk)-\epsilon_F\right)$. For metals we expect
        $g(\epsilon_F)$ to be finite, while it should go to $0$ for
        insulators. We now consider how $g(\epsilon_F)$ behaves with
        carrier density for different values of $\Delta$. These
        $\Delta$ values
        are not obtained from a self-consistent solution of the gap
        equation; rather our intention is to vary $\Delta$ and
        understand the range of values for which we can recover the
        metallic behaviour seen in the experiments. For each $\Delta$, we
        solve the number equation to obtain the Fermi level for the
        excitonic quasiparticles.  In
        Fig.~\ref{Fig4}(a), we plot $g^0/g(\epsilon_F)$ as a function of
       	the order parameter $\Delta$ at CNP. Here $g^0$ is the density of
        states at the Fermi level of the non-interacting system at
        CNP. For
        $\Delta>1.63 ~\textrm{meV}$ the system is insulating in nature and below that it's a metal. Since the
        experimental data~\cite{anindya_thermo} clearly shows metallic behaviour at low
        temperatures for all densities near CNP, this provides an
        upper bound for the possible values of $\Delta$. Within the
        mean field theory, this in turn provides an upper bound on the
        interaction strength $u_0$. For $\Delta_{CNP} < 1.63
        ~\textrm{meV}$, we obtain $u_0 < 13.6 ~\textrm{meV}$. In this work, we use a value of $u_0=10.9\;\tr{meV}$ which is consistent with earlier work~\cite{Cea2020}.

        \begin{figure*}[t!]
	\includegraphics[width=\textwidth]{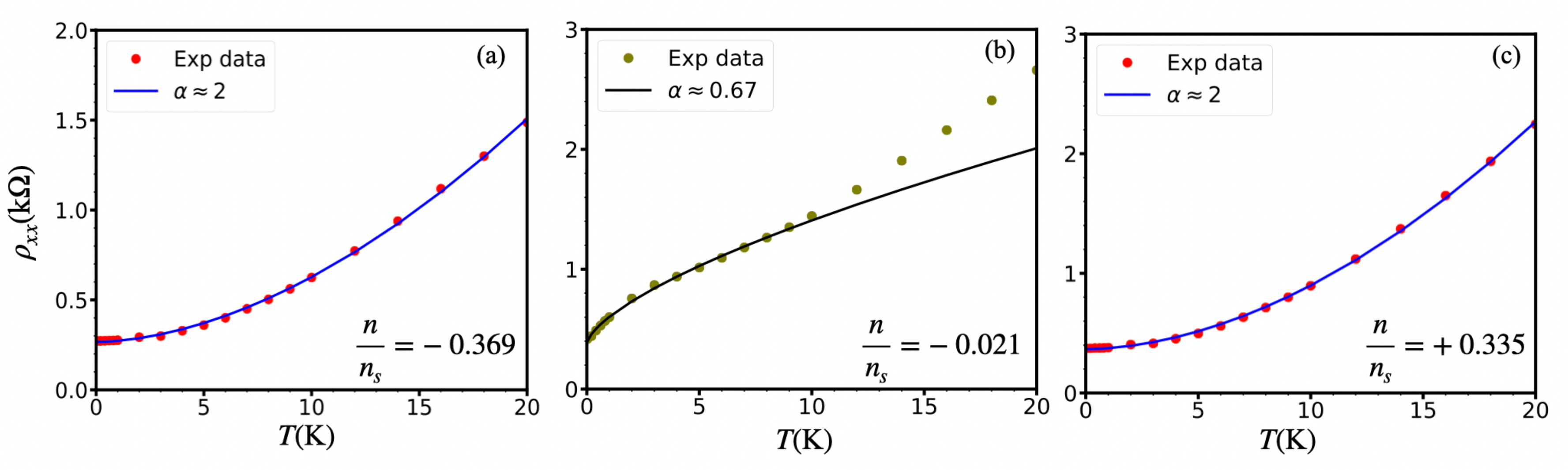}
	\caption{(a)-(c) shows the experimental data for the
          temperature dependence of longitudinal resistivity near and
          away from CNP. Note that the non-interacting Fermi
          temperature $T_F \sim 23 K$. In (a) and (c) a fit to the
          data of the form $R_0+A\;T^\alpha$ gives $\alpha \sim 2$,
        showing Fermi liquid scaling. The data near CNP in (b) clearly shows
          sublinear behaviour of the resistivity below 10K and an
          almost linear behaviour above it. Here the data up to $10$K
          is fitted to $R_0+A\;T^\alpha$, yielding $\alpha
          =0.67$. This sublinear temperature dependence is a
          manifestation of an underlying non-Fermi liquid in the system.}
	\label{Fig:Fig_5}
\end{figure*}

         In recent experiments, resistance of tDBLG at low
         temperatures have shown a distinct two-peak structure as a
         function of carrier density near CNP. This is seen in the
         region where electron and hole pockets are simultaneously
         present. A representative experimental data is shown in
         Fig.~\ref{Fig4}(b). We will use the inverse of the density of
         states at the Fermi level as a proxy for the resistivity of
         the system. This assumes that Fermi velocity is not sharply
         anisotropic in the density range considered, which is valid 
         for tDBLG near CNP with its small electron and hole
         pockets. In Fig.~\ref{Fig4}(c), we plot $g^{-1}(\epsilon_F)$ as a function of carrier density
         for a system with indirect excitonic condensate
         ($u_0=10.9~\textrm{meV}$). We clearly see the presence of two peaks,
         as seen in the experiments. For comparison, we have also
         plotted $g^{-1}(\epsilon_F)$ for the non-interacting system,
         which does not show any structure in the relevant density
         range. This shows that the presence of excitonic condensate
         can explain the double peak structure seen in experiments. In Fig.~\ref{Fig4}(d) we plot $g^{-1}(\epsilon_F)$ with carrier density for different interaction strengths and see that the two-peak feature is lost for $u_0<10 \;\tr{meV}$. This is close to the lower bound obtained from experimental $T_c$ discussed in the earlier section. 

	In order to theoretically understand the origin of the two
        peaks, we look at the dispersion of the $6$ bands in presence
        of the excitonic condensate (detail band dispersion of all 6 bands are plotted in Appendix.~\ref{appB}). In Fig.~\ref{Fig4}(e-h), we plot the Fermi surfaces of the quasiparticles (with spectrum
        $E^\pm_{\QQ_i}(\kk)$) in the mBZ for four
        different carrier densities, $n/n_s=0,~0.03,~0.06,~0.10$. At
        CNP ( Fig.~\ref{Fig4}(e) ), we find that both $E^+$ and $E^-$ bands pass through the
        Fermi level, and hence give rise to their respective Fermi
        surfaces (shown by green and pink lines respectively). This
        leads to the finite density of states at Fermi energy and
        hence to metallic behaviour in the system. As we move away
        from the CNP on the electron doped side, the Fermi surfaces
        corresponding to $E^-$ reduces, while that of $E^+$ bands
        increases in size ( Fig.~\ref{Fig4}(e)-(h)). This leads to a reduction in the density of
        states at the Fermi level and hence to an increase in
        resistivity till the $E^-$ Fermi surface vanishes at
        $n/n_s=0.06$ ( Fig.~\ref{Fig4}(g) ), which corresponds to the
        peak in the inverse density of states. Beyond this point, we
        see additional Fermi surfaces as the chemical potential enter
        one of the bands which was gapped at CNP. This leads to a
        large increase in the density of states at the Fermi level and
        consequently a sharp decrease in the resistivity, explaining
        the peak seen in the resistivity for $n>0$. A similar
        argument, with the roles of $E^+$ and $E^-$ reversed, explains
        the peak for $n<0$. We would like to note that the transport
        features are expected to be more smeared than single particle
        features due to the presence of inhomogeneities in the
        system. This is consistent with the experiments, which see
        relatively smoother features compared to the theoretical
        estimates.

         \section{Temperature Dependence of Resistivity and Non-Fermi
           Liquid Behaviour\label{nfl}}

         An intriguing signature of strong electronic correlations is seen
   in recent experiments~\cite{anindya_thermo} on the temperature dependence of resistance
   of tDBLG near the charge neutrality point. The resistance shows a sublinear
   ($R=R_0+A\;T^\alpha$ with $\alpha\approx2/3$) behaviour up to $T\sim 10~K$, and an almost linear
   behaviour above that temperature(Fig.~\ref{Fig:Fig_5}(b)). In contrast, the temperature
   dependence of resistance at larger electron (hole) densities,
   ($n/n_s = +0.335$) and ($n/n_s = -0.369$), are superlinear and can be
   fitted to the standard quadratic Fermi
   liquid scaling, $R=R_0+A\;T^\alpha$ with $\alpha\approx2$, expected for a 2d system with small
   Fermi pockets,  both from electron-electron
   and electron-phonon scattering. This is shown in Fig~\ref{Fig:Fig_5}(a) and (c).

   This novel sublinear
   behaviour cannot be explained within a Fermi liquid paradigm of
   perturbative effects of electronic interactions around a free Fermi
   gas. Electron-electron interactions lead to a $\sim T^2$ dependence
   of the resistivity. In 2 dimensions, the density of states of
   longitudinal phonons  $\rho(\omega) \sim \omega$. This leads to
   $\sim T^4$ dependence of resistivity when small angle scattering
   dominates at low temperatures and $\sim T^2$ behaviour of
   resistivity when scattering at all angles contributes. For
   compensated semi-metals like tDBLG, with small Fermi surfaces, one
   would expect the resistivity $\sim T^2$. At higher temperatures
   (beyond the Bloch Gruneissen temperature), the scattering from
   classical phonons would lead to a linear $T$ dependence of
   resistivity~\cite{phonon_SDS}. None of these mechanisms can explain a sublinear
   temperature dependence of resistivity, as is seen in the experiments.

   In this section, we will show that a tDBLG system near CNP, which is at the
   precipice of a quantum criticality towards the formation of an
   excitonic condensate, will show a sublinear $T^{2/3}$ temperature
   dependence of resistance. At criticality, the low energy gapless
   fluctuations of the excitonic (CDW) order parameter would be
   strongly Landau damped due to the presence of the electronic states of
   the metal around the Fermi surfaces. The scattering between
   electrons and the Landau damped fluctuations lead to a non-analytic
   low energy decay rate for the single particle electronic excitations, resulting in non-Fermi liquid behaviour. For compensated semimetals like
   tDBLG, with small Fermi surfaces, the transport scattering rate and
   the quasiparticle scattering rate has the same energy dependence. This
   non-analytic frequency dependence of the scattering rate leads to
   the sublinear temperature dependence of resistivity in this
   system. We note that this mechanism is an alternate to theories of
   Planckian metal~\cite{Aavishkar_planck}, which predicts linear
   temperature dependence of resistivity all the way down to $T=0$.

   The fluctuations of the order parameter, $\phi(q,\omega)$ are
   governed by an action,
   \beq
   S_{fl} = \int d^2 q~ \int d\omega ~\phi^\ast(q,\omega)
   \Pi^{-1}_{inter}(q,\omega) \phi(-q,-\omega)
   \eeq
where $\Pi_{inter}$ is the (interacting) inter-band polarizability of
the system. When driven to the critical point, $
\Pi^{-1}_{inter}(Q,0)=0$; i.e. the gap vanishes (this can be seen within
a simple RPA like theory), and generically we would have

\beq
   \Pi^{-1}_{inter} \sim 
   \left[\omega^2-c^2\bar{q}^2-i \gamma \frac{|\omega|}{|\bar{q}|}\right]
   \eeq
   where $\bar{q}=q-Q$, $c$ and $\gamma$ are constants denoting the speed of the
   fluctuation waves and the scale of damping. The singular
   $\left\vert \frac{\omega}{\bar{q}}\right\vert$ damping is a consequence of the presence
   of low energy fermions, and derives from the imaginary part of the
   non-interacting inter-band polarizability.

\begin{figure}[t]
	\includegraphics[width=\columnwidth]{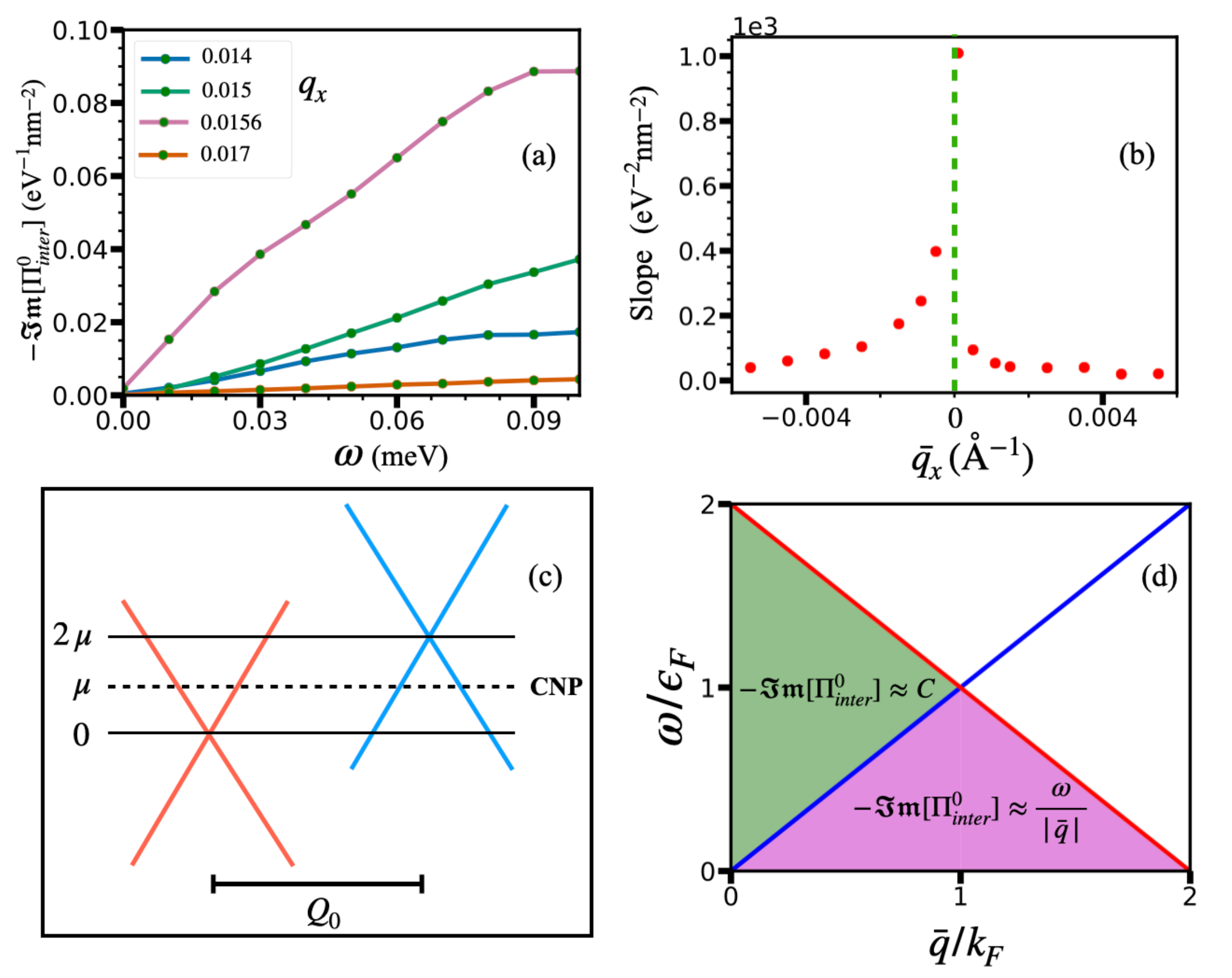}
	\caption{(a) Imaginary part of the interband polarizability
          function is plotted with frequency for different $q_x$
          values. Note that the direction of the $\vec{q}$ variation
          has been taken along the $[1,0]$ direction, and around
          $Q_1\approx(0.016,0)$, which is the connecting wavevector
          between one of the e-h pockets. The low frequency behaviour
          is linear with the slope 
          showing a huge increase near $Q_1$. (b) The slope at
          different $q_x$ from (a) is plotted here to show the
          divergence at $\vec{q}=Q_1$ (or, $\bar{q}=0$). (c) A simple
          theoretical model of the electron and hole pockets for
          calculating the inter-band polarizability of the system is
          shown here. The conduction (blue) and valence (red) Dirac
          bands are separated by $Q_0$ momentum inverted with respect
          to each other. The Dirac points are separated in energy by
          $\epsilon_0$. At CNP,  the chemical potential
          $\mu=\epsilon_0/2$. (d) Schematic showing the regions in
          $\bar{q}-\omega$ plane where the imaginary part of the
          polarizability of the simple model shows (i) singular Landau
          damping
          $\mathfrak{Im}(\Pi^0_{inter})\approx\frac{\omega}{|\bar{q}|}$
          and (ii) a constant damping rate. The non-Fermi liquid
          behaviour resulting from the Landau damping leads to the
          $T^{2/3}$ dependence of the resistivity.}
	\label{Fig:Fig_6}
\end{figure}

   \begin{widetext}
   The non-interacting interband polarizability of the system is given by
   \beq
   \Pi_{inter}^0(q,\omega) = \sum_k \frac{
     f[\epsilon_c(k)]-f[\epsilon_v(k+q)]}{\omega+ i\eta
     +\epsilon_c(k)-\epsilon_v(k+q)} |\langle
   \psi_c(k)|\psi_v(k+q)\rangle|^2 ~+~ c\leftrightarrow v
   \label{intebandpol}
   \eeq
   where $f$ is the Fermi function and $|\psi_{c(v)}(k)\rangle$ is the
   Bloch wavefunction of the corresponding band in the mBZ. To see the  Landau damping, in Fig.~\ref{Fig:Fig_6}(a), we plot the imaginary
   part of $\Pi^0_{inter}$ as a function of $\omega$ for several values
   of $q$ along the $[1,0]$ direction. Note that $Q_1\sim[0.016,0] A^{-1}$ is along
   $[1,0]$ direction, so we will be crossing the exciton wave-vector
   in the process. At low $\omega$, the plots are linear and we extract
   a slope from this linear part of the graph. The slope is plotted as
   a function of $q$ in Fig.~\ref{Fig:Fig_6}(b). The slope diverges at
   $q=Q_1$, showing the singular nature of the Landau damping. We have
   checked by taking momentum cuts along other directions passing
   through $Q_1$, that the divergence of the slope happens whenever we
   approach $Q_1$ along any direction, i.e. there is no difference
   in the scaling between directions along and perpendicular to the
   exciton/CDW wave-vector (see Appendix.~\ref{appA} for details).
 \end{widetext}
 
   We can understand this further by constructing a simple model for
   the valence and conduction band. The electron pocket of the conduction band is modeled by a
   Dirac dispersion centered around $k=0$, i.e. $\epsilon_c(k)= v_F|k|$,
   while the hole pocket in the valence band is modeled by an inverted Dirac dispersion
   centered around $Q$, i.e. $\epsilon_v(k) = \epsilon_0-
   v_F|k-Q|$ in Fig.~\ref{Fig:Fig_6}(c). Here $\epsilon_0$ is the energy difference between the
   Dirac points. At CNP, the chemical potential sits at
   $\mu=\epsilon_0/2$ giving a Fermi wave vector $k_F=\mu/v_F$ for electrons and holes. Usually, the orthogonality of the band wavefunctions as $q\rightarrow0$ plays an important role in determining the low $\omega,q$ behaviour of $\Pi^0_{inter}$ in graphene. However, here we are interested in $\Pi^0_{inter}$ near $q=Q$ where orthogonality considerations do not play a role. The band overlaps do not change the scaling of
   various terms (although they can change the value of the coefficients like
   $\gamma$ etc.). Hence we ignore band overlaps in our calculation of $\Pi^0_{inter}$. In this case, the polarization function can be
   calculated exactly and the detailed formulae are given in
   Appendix~\ref{appA}.  Near $Q$, the imaginary part of $\Pi^0_{inter}$ has the
   following form:\\
   \begin{widetext}
    \bqa
   -\mathfrak{Im}~\Pi^0_{inter}(q,\omega) & \sim \frac{k_F^3}{4\pi\mu^2} \left\vert\frac{\omega}{\bar{q}}\right\vert & ;
   (\omega<v_F\bar{q} \,\,\textrm{when}\,\, \bar{q}<k_F) \,\textrm{or,}\,(\omega<\epsilon_0-v_F\bar{q} \,\,\textrm{when}\,\, \bar{q}>k_F)\\
   \nonumber  & \sim \frac{k_F^2}{8\mu} ~~~~~~~& ;
   ~~ \epsilon_0-v_F\bar{q}>\omega > v_F\bar{q}
   \eqa
   \end{widetext}
  
Thus both the numerical calculation on the full model and the analytic
calculation with the simple model shows the presence of the singular
Landau damping around $q\sim Q$.
The Landau damped fluctuations have an energy momentum relation $\omega
\sim \bar{q}^3$, leading to a density of states $\rho(\omega) \sim \omega
^{-1/3}$. The scattering of electrons by these fluctuations lead to a
single particle scattering rate
\begin{eqnarray}
	\Gamma_{qp} &\propto&\int_{0}^{\omega}d\omega'
        ~\rho(\omega') \sim \omega^{2/3} ,\tr{at}\,T=0\\
        &\propto&\int_{0}^{\omega_c}d\omega
        \,\rho(\omega)\, coth(\frac{\omega}{2T})\sim T^{2/3} ,\tr{at}\,T\neq0
\end{eqnarray}
   where $\omega_c $ is a cutoff  below which the long wave-length
   description mentioned above holds with the condition $\omega_c>> T$. The damping rate $\omega^{2/3}$ is larger than the energy of excitations leading to non-Fermi liquid behaviour. Since the Fermi pockets in tDBLG
   near CNP are small, scattering at all angles contribute to
   transport and hence the transport scattering rate $\Gamma_{tr}\sim\Gamma_{qp} \sim
   T^{2/3}$. This is the origin of the $T^{2/3}$ dependence of
   resistance in the system. Thus the sublinear temperature dependence
   is a reflection of an underlying non-Fermi liquid state in a
   quantum critical metal.

   Starting with the work of Hertz and Millis,~\cite{Hertz_qcp, millis_qcp}, non-Fermi liquid behaviour of itinerant quantum critical systems
   have been studied in the context of high temperature
   superconductivity~\cite{Polchinski,Senthil2004,Sachdev2000}, for the
   antiferromagnetic metal~\cite{Sachdev_2012,Vekhter2004,liu_qcp_exp} and for the nematic
   transition in metals~\cite{Metlitski2010,Klein2020} using sophisticated
   renormalization group techniques. The general conclusion is that
   the imaginary part of the electron self energy $\sim \omega^{2/3}$
   when the order parameter is at $0$ wave
   vector~\cite{Metlitski2010} (as in nematic transition), while we
   get a $\sim \omega^{1/2}$ dependence for a finite $Q$ wave-vector
   order  like spin density waves~\cite{Metlitski2}. We note that in our case the finite
   wave-vector relates the center of the electron pocket to the center
   of the hole pocket, so the electron and hole Fermi surfaces lie on
   top of each other when shifted by this wave vector. Thus, although
   the CDW is formed at finite wave vector, our
   itinerant quantum criticality is similar to the nematic case with
   effective $q=0$ and we recover the $\sim \omega^{2/3}$ scaling.

   In the experiments, it is seen that the resistivity shows a linear
   temperature dependence above $T=10K$. We note that the Fermi
   temperature corresponding to the e-h pockets is $ T_F \sim 20 - 25  K$ and
   one would expect quantum behaviours to vanish beyond this
   temperature.  For a 2-dimensional system with small Fermi surfaces,
   a more relevant scale would be the effective ``Bloch
   Gruneissen'' temperature of these critical fluctuations, i.e. their
   energy for $\bar{q} \sim 2k_F$. While this requires interaction
   renormalized estimates of $\gamma$ and $c$, which are  beyond the
   scope of this paper, one would expect this temperature to
   be lower than $T_F$. If we assume that beyond $T=10K$, these
   fluctuations are classical, we can obtain the linear temperature
   dependence of resistivity in this regime.

   As we move away from the CNP on either side, the mismatch between
   the electron and hole Fermi surfaces increases, and the Landau
   damping would be shifted to finite $\bar{q} > \Delta k_F$, where
   $\Delta k_F$ is the mismatch of the Fermi wave vectors of the electron and hole pockets. In this
   case, the low temperature behaviour of resistivity would deviate
   from the $T^{2/3}$ scaling. Assuming the $|\omega/\bar{q}|$ scaling
   of the Landau damping is cut-off on the scale of $\Delta k_F$,
   i.e. it scales as  $|\omega/\Delta k_F|$, one would get a linear
   $T$ dependence of the resistivity. However, this exponent would be
   strongly influenced by disorder and Fermi surface anisotropies. In
   general, one would expect the exponent to increase as one goes away
   from CNP. Once the simultaneous presence of electron and hole
   pockets vanish at larger doping, one gets back the standard $T^2$
   scaling of the Fermi liquid theory.

   \section{Conclusion\label{concl}}
	Strong electronic interactions determine the plethora of symmetry broken phases in magic angle tBLG and tTLG. In contrast, tDBLG is usually thought of as a plain vanilla metal in the absense of electric/magnetic fields. The metallicity of magic angle tDBLG near CNP comes from the overlap of flat conduction and valence bands which creates small electron and hole pockets at the Fermi surface.\\
	In this paper, we have considered the possible effects of Coulomb interaction on these small e-h pockets in the compensated semi-metal near CNP. We show that the interactions lead to the formation of indirect excitons (i.e. particle-hole pairs with finite momenta). The condensation of such pairs lead to the formation of CDW states. We show that for a reasonable range of interaction parameters, this ordered state has reorganised Fermi pockets and hence metallic behaviour is expected. However, the density of electronic states is strongly renormalised in the process. The inverse density of states at Fermi level shows peaks at finite doping on either side of CNP. This can explain the peaks in the resistance as a function of density seen in recent low temperature experiments~\cite{anindya_thermo}. We further show that the Landau damped critical fluctuations of the excitonic order can give rise to non-Fermi liquid behaviour of the electrons with a scattering rate, $\Gamma\sim T^{2/3}$. For systems like tDBLG, with small Fermi pockets, this can give rise to sub-linear $T^{2/3}$ resistance seen in recent experiments~\cite{anindya_thermo}.\\
	Our theoretical predictions thus strongly indicate that effect of excitonic/CDW order and its fluctuations have already been observed through the non-Fermi liquid temperature dependence of resistivity. A hallmark of CDW insulators is the current produced by sliding mode when the order is depinned by finite energy probes(non-linear/AC conductivity). However, in a metallic system like tDBLG, these contributions would be masked by the usual single particle contributions. We believe the fluctuations of the finite particle-hole condensate should give additional contributions to current noise but we leave this calculation for a future manuscript. 
\section{Acknowledgement}
R.S. would like to thank Mohit Randeria for useful discussions. UG and RS acknowledge computational facilities at the Department of Theoretical Physics, TIFR Mumbai.  R.S. acknowledges support of the Department of Atomic Energy, Government of India, under Project Identification No. RTI 4002. UG would like to thank Md. Mursalin Islam for useful discussions related to computational part of the project.

\bibliography{Exciton_paper}

	\newpage
%	\begin{widetext}
	\onecolumngrid

	\section*{Appendix}
	\begin{appendices}
	
	\section{Interband Polarizability and Landau damping}\label{appA}
	We are interested in the low energy dispersion of valence and conduction bands in tDBLG, especially the modes which constitute the electron ad hole pockets near the CNP. For these modes, the energy dispersion can be approximated as two Dirac cones separated in momentum and energy as shown in Fig.~\ref{Fig:Fig_6}(c). We use the hole band of the Dirac cone inverted around higher Dirac point and the electron band around lower Dirac point as,
	\begin{eqnarray}
		\epsilon_\kk^c&=&v_F|\kk| \nonumber \\
		\epsilon_\kk^v&=&2\,\mu-v_F|\kk-\QQ_0|
	\end{eqnarray}
	Here, $	\epsilon_k^{c(v)}$ is the energy dispersion of conduction(valence) band, $\mu$  is the chemical potential at CNP, $v_F$ denotes the Fermi velocity of the effective Dirac points and the momentum separation wavevector between two Dirac cones is given by $\QQ_0$. Also one can note that exciton order parameter $\Delta(\qq)\sim \sum_\kk \langle C^{\dagger c}_{\kk}
	C^v_{\kk+\qq}\rangle$, so its fluctuations, $\chi_{\Delta\Delta}(q,t-t')=i\theta(t-t')\sum_{\kk\kk'}\braket{[C^{c\dagger}_\kk(t)C^{v}_{\kk+\qq}(t),C^{v\dagger}_{\kk'}(t')C^{c}_{\kk'-\qq}(t')}]>$ are related to the non-interacting interband polarizability,\\

		\begin{eqnarray}
			\Pi^0_{cv}(\qq,\omega)=\sum_{\kk\in \textrm{mBZ}} \frac{f(\epsilon_\kk^c-\mu)-f(\epsilon_{\kk+\qq}^v-\mu)}{\omega+\epsilon_\kk^c-\epsilon_{\kk+\qq}^v} |\braket{\psi_\kk^c\:|\:\psi_{\kk+\qq}^v}|^2 \label{lindeq1}
		\end{eqnarray}
			Where the total polarizability $\Pi^0_{inter}=\Pi^0_{cv}+\Pi^0_{vc}$. Usually the orthogonality of the band wavefunctions as $q\rightarrow0$ plays an important role in determining the low $\omega,q$ behaviour of $\Pi^0_{inter}$ in Graphene. However, here we are interested in $\Pi^0_{inter}$ near $q=Q$ where orthogonality considerations do not play a role. The band overlaps do not change the scaling of
			various terms. Hence we can drop the term $|\braket{\psi_\kk^c\:|\:\psi_{\kk+\qq}^v}|^2$ from Eq.~\ref{lindeq1} and write it as,
		\begin{eqnarray}
			\Pi^0_{cv}(\qq,\omega)\sim\int \frac{ d^2\kk}{4\pi^2} \frac{f(\epsilon_\kk^c-\mu)} {\omega+\epsilon_{\kk-\bar{\qq}}^c+\epsilon_{\kk}^c-2\mu}-\int \frac{ d^2\kk}{4\pi^2}  \frac{1-f(\epsilon_\kk^c-\mu)}{\omega+\epsilon_\kk^c+\epsilon_{\kk+\bar{\qq}}^c-2\mu}
			\label{lindeq2}
		\end{eqnarray}
		where $\bar{\qq}=\qq-\QQ_0$. Now using dimensionless parameters, $X=k/k_F,\; Y=\bar{q}/k_F, \textrm{and}\, Z=\omega/\mu$, the Eq.~\ref{lindeq2} can be rewritten as,

		\begin{eqnarray*}
			\Pi^0_{cv}(\qq,\omega)\sim-\frac{k_F^2}{4\pi^2\mu}[-\int_0^1 XdX\int d\phi \frac{1} {Z-2+X+\sqrt{X^2+Y^2-2XY\cos{\phi}}}\\
			+\int_1^\Lambda XdX\int d\phi \frac{1} {Z-2+X+\sqrt{X^2+Y^2+2XY\cos{\phi}}}]
		\end{eqnarray*}

	Here $\phi$ is the azimuthal angle between $\kk$ and $\bar{\qq}$. Here $\Lambda$ is an ultraviolet cut-off. We note that the Landau damping we calculate is a low energy property, which is independent of $\Lambda$. We are interested in the imaginary part of the polrizability, $\Pi^{''}\equiv\mathfrak{Im}[\Pi^0_{cv}]$, which then becomes,

		\begin{eqnarray*}
			\Pi^{''}\sim-\frac{k_F^2}{4\pi\mu}[\int_0^1 XdX\int_{-1}^{+1} \frac{du}{\sqrt{1-u^2}} \delta[Z-2+X+\sqrt{X^2+Y^2-2XYu}]\\
			-\int_1^\Lambda XdX\int_{-1}^{+1} \frac{du}{\sqrt{1-u^2}} \delta[Z-2+X+\sqrt{X^2+Y^2+2XYu}]]
		\end{eqnarray*}

	Here $u=\cos{\phi}$. This azimuthal integral can be done analytically to get,

		\begin{eqnarray*}
			\Pi^{''}\sim-\frac{k_F^2}{4\pi\mu}[\int_0^1- \int_1^\Lambda] XdX\frac{\; \theta [4X^2Y^2-(X^2+Y^2-(2-Z-X)^2)^2]}{\sqrt{4X^2Y^2-(X^2+Y^2-(2-Z-X)^2)^2}}
		\end{eqnarray*}

	which can also be written as,

		\begin{eqnarray}
			\Pi^{''}\sim-\frac{k_F^2}{4\pi\mu}[\int_0^1- \int_1^\Lambda] XdX\frac{\; \theta [\{(2-Z)^2-Y^2\}(X^+-X)(X-X^-)]}{\sqrt{4\{(2-Z)^2-Y^2\}(X^+-X)(X-X^-)}}
		\end{eqnarray}

	where, $X^\pm=((2-Z)\pm Y)/2$. We will work with $Y, \;Z>0$ and $Y,Z<<1$ so that, $Z<2$ and, $Y<2-Z$. In this case, $X^+>X^->0$. Now depending on the value of $Y,\; Z$ we can have different regions in the phase space where the integrals take qualitatively different forms,\\
	\\
	\textbf{Case-(i)}:\\
	$0<Z<1$ and $0<Y<Z$, \\or, $1<Z<2$ and $0<Y<2-Z$,
	\begin{eqnarray}
		-\Pi^{''}&\sim&\frac{k_F^2\;(2-Z)}{16 \mu \sqrt{(2-Z)^2-Y^2}}\nonumber\\
		&\approx&\frac{k_F^2}{16\mu} ;\;(Y<<2-Z)
	\end{eqnarray}
	\\
	\textbf{Case-(ii)}:\\
	$0<Z<1$ and $Z<Y<2-Z$,

		\begin{eqnarray}
			-\Pi^{''}&\sim&\frac{k_F^2}{8\pi \mu\sqrt{(2-Z)^2-Y^2}}[(2-Z)\;\sin^{-1}(\frac{Z}{Y})-\sqrt{Y^2-Z^2}]\nonumber\\
			&\approx&\frac{k_F^2}{8\pi \mu}\frac{Z}{Y};\;(Y\rightarrow0 , Z/Y\rightarrow0)\nonumber\\
			&\sim&\frac{k_F^3}{8\pi \mu^2}\frac{|\omega|}{|\bar{q}|}
		\end{eqnarray}
	Therefore we analytically show the Landau damping factor($\frac{|\omega|}{|\bar{q}|}$) arising from imaginary part of the interband polarizability function assuming two simple Dirac bands separated from each other in energy and momentum space. One can easily see that $\Pi^0_{vc}$ will give the same contribution. Collecting all these,
	\begin{eqnarray}
		-\mathfrak{Im}~\Pi^0_{inter}(q,\omega)&\sim&\frac{k_F^2}{8\mu};\,\,\, Y<Z<2-Y \,\textrm{for}\,Y<1\\
		&\sim&\frac{k_F^3}{4\pi \mu^2}\frac{|\omega|}{|\bar{q}|};\,\,\, (Z<Y \,\textrm{for}\, Y<1) \,\textrm{or,}\,(Z<2-Y \,\textrm{for}\, Y>1)
	\end{eqnarray}\\
%	\end{widetext}
	\begin{figure}[H]
		\includegraphics[width=\textwidth]{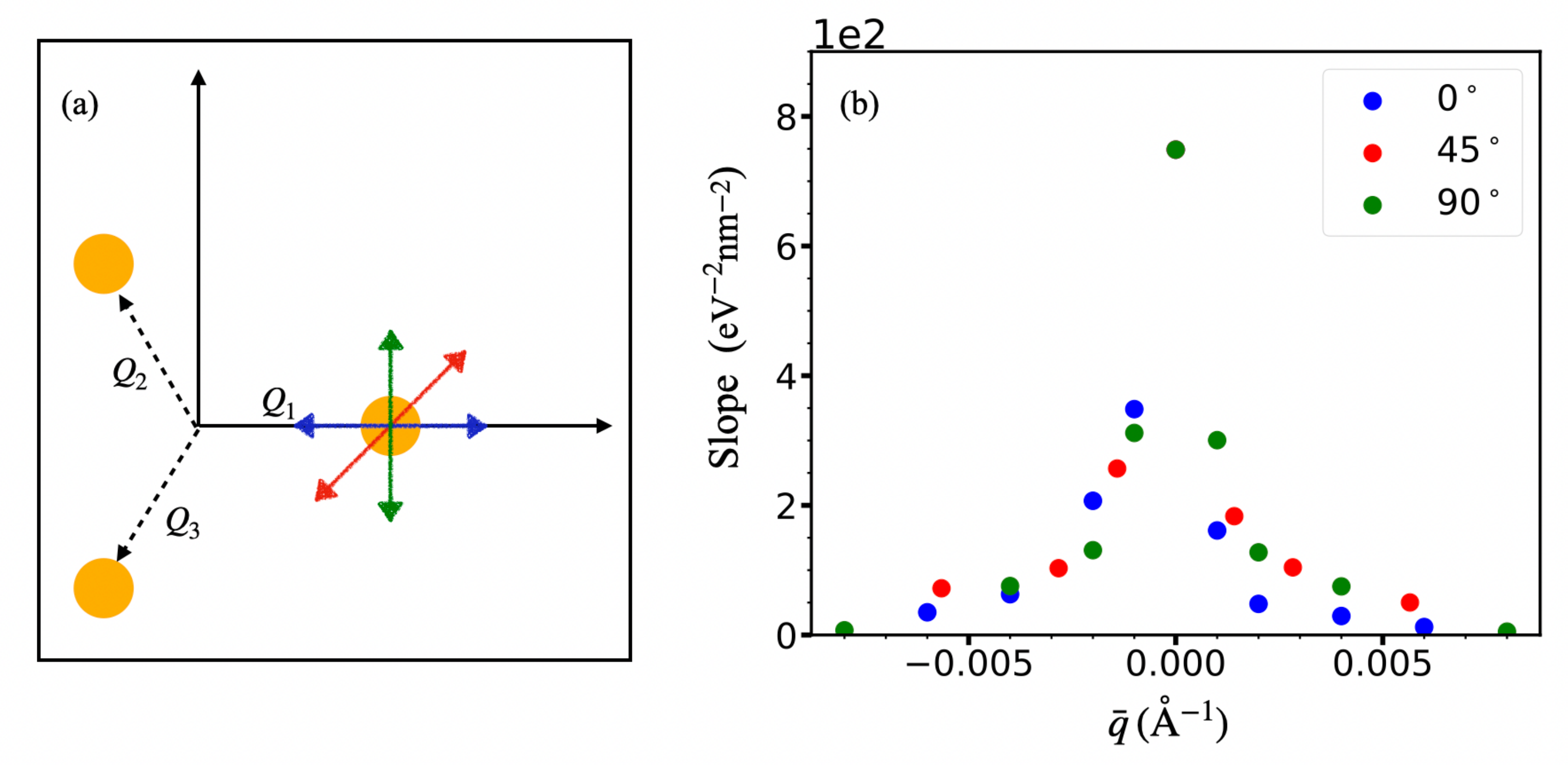}
		\caption{In (a), we have shown a schematic of the three $C_3$ symmetric critical points in the momentum space. Around the $Q_1$ point, we have taken the $q$-cuts of $\frac{d\,\Pi^{''}}{d\omega}|_{\omega\rightarrow0}$ along the directions drawn and plotted them in (b). Notice that $q$ has been scaled to $\bar{q}=q-Q_1$. The slope along all three directions [1,0](Blue),[1,1](Red) and [0,1](Green) show divergence at the critical point and a very small anisotrpy on top of the isotropic background. One thing to note is for this figure (b) we have used a different set of parameters (momentum grid $350\times350$ and broadening $\eta=0.05\,\text{meV}$) than for Fig.~\ref{Fig:Fig_6}(b) where we used (momentum grid $800\times800$ and broadening $\eta=0.02\,\text{meV}$). That is the reason the divergence scale of the slopes are a different for these two figures which is expeceted.)}
		\label{appnd_fig1}
	\end{figure}
	As shown in the main text, the numerical evaluation of $\Pi^{''}$ using the actual tDBLG dispersion and wavefunctions also give rise to similar singular damping. In the main text, we looked at $\frac{d\,\Pi^{''}(q,\omega)}{d\omega}|_{\omega\rightarrow0}$ as a function of $q$. Here in Fig.~\ref{appnd_fig1} (b), we additionally plot the slope $\frac{d\,\Pi^{''}}{d\omega}|_{\omega\rightarrow0}$ along a cut in $q$ which passes through $Q_0$ and moves along [0,1]($90^\circ$) and [1,1]($45^\circ$) directions (the directions are schematically shown in Fig.~\ref{appnd_fig1}(a)). We see that the slope diverges at $Q_0$ irrespective of the direction in which it is approached. Thus the Landau damping has only small anisotropies riding on an isotropic background, even though the Fermi velocities at the Dirac points are anisotropic. This isotropy is important for scaling arguments used to explain $T^{2/3}$ resistivity.	
	\section{ Band renormalization due to exciton formation}\label{appB}
\begin{figure}[h!]
	\includegraphics[width=16cm]{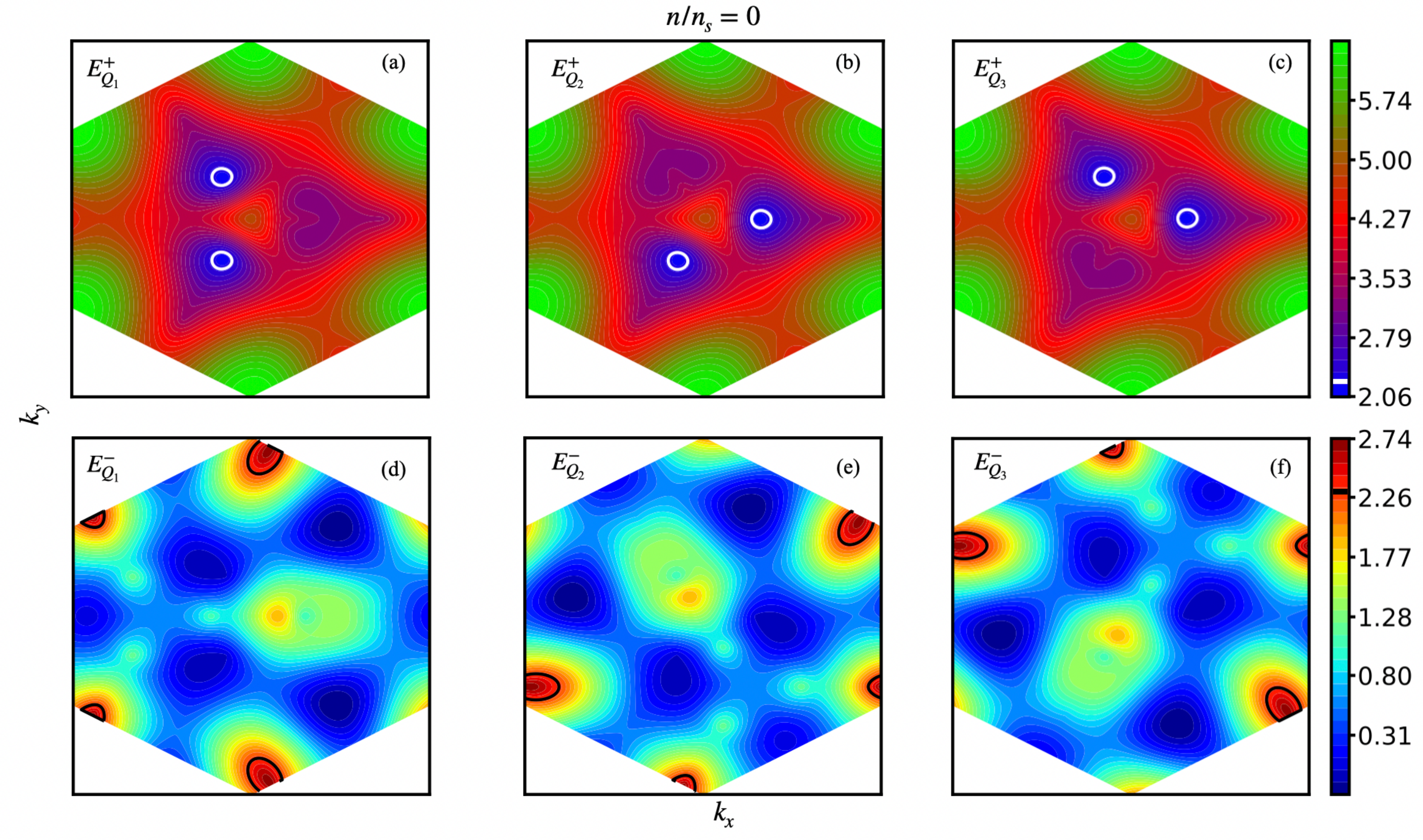}
	\caption{In (a)-(c) we show color plots of the renormalized conduction bands due to the presence of excitonic order parameters for the three e-h pocket connecting wave-vectors $Q_{1(2)(3)}$ respectively at CNP. We plot the renormalized valence bands at CNP in (d)-(f). The solid lines represent the Fermi surface at CNP. Here both the electron and hole pockets contribute to the resistivity of the system.}
	\label{appnd_fig2}
\end{figure}

\begin{figure}[h!]
	\includegraphics[width=16cm]{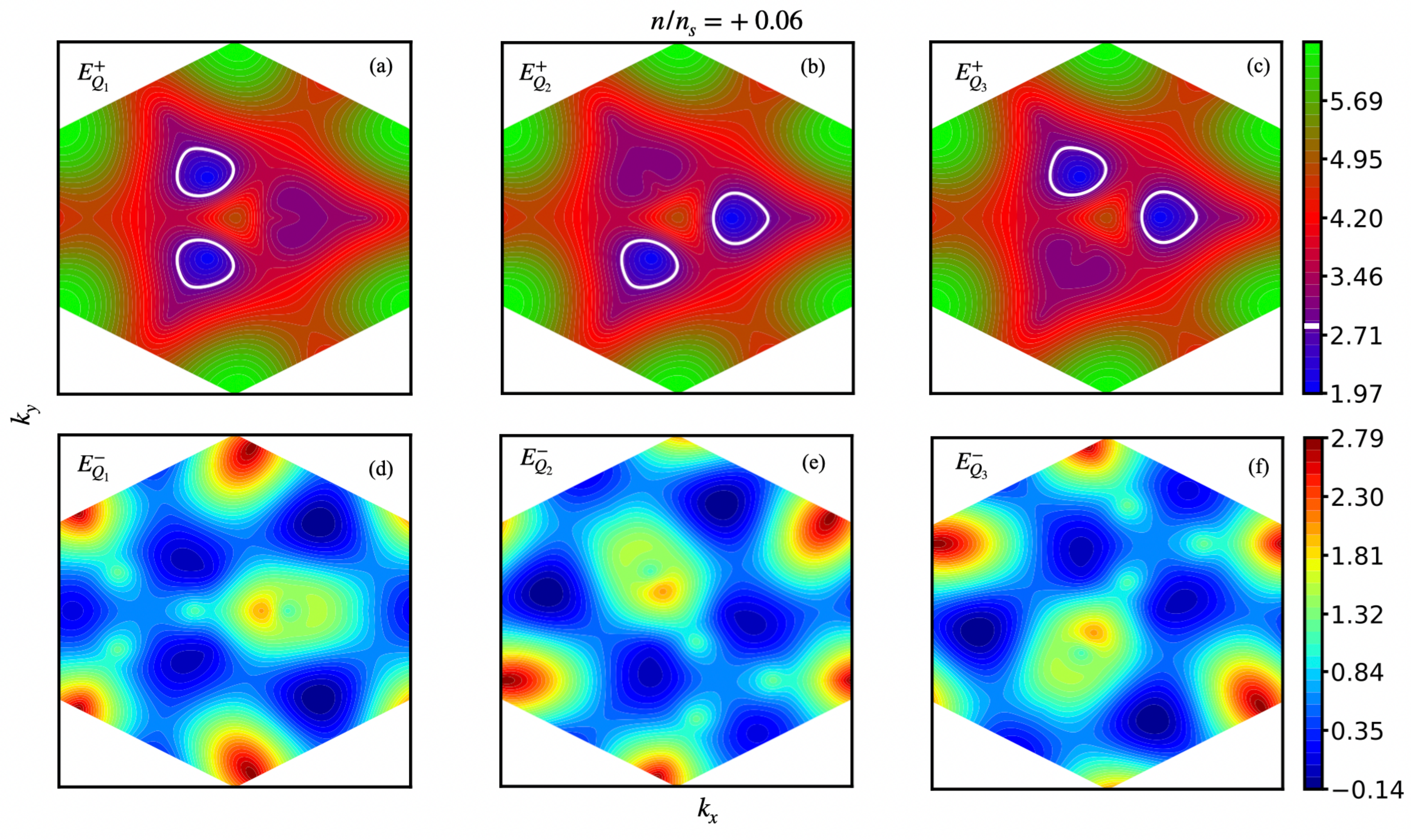}
	\caption{Similar to Fig.~\ref{appnd_fig2} here we have plotted the exciton bands for density $n/n_s=0.06$. One can notice that the hole pockets have vanished which explains the suppression of density of states which leads to increase in the resistivity.}
	\label{appnd_fig3}
\end{figure}

\begin{figure}[h!]
	\centering
	\includegraphics[width=16cm]{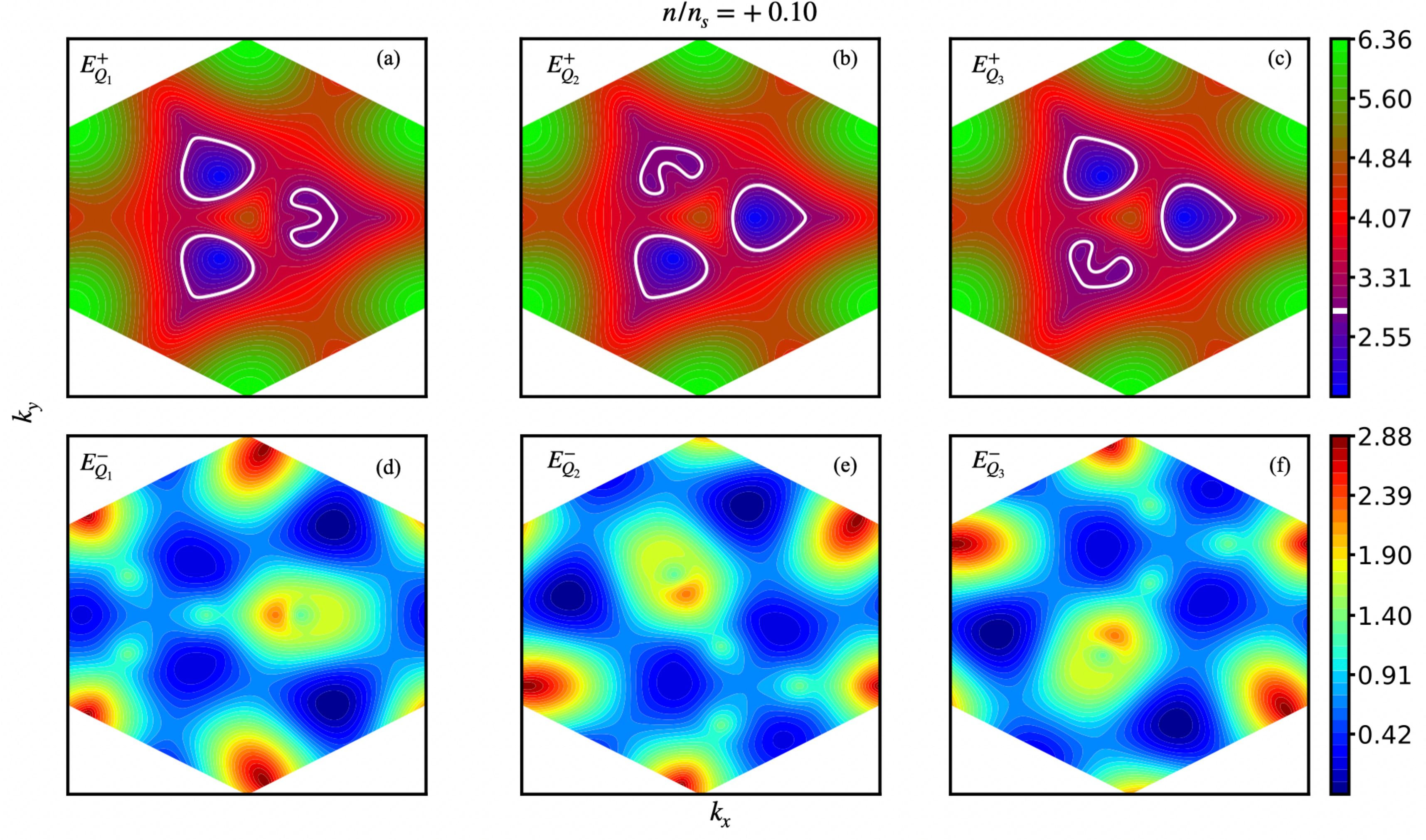}
	\caption{Similar to Fig.~\ref{appnd_fig2} here we have plotted the exciton bands for density $n/n_s=0.10$. Here one can notice that extra electron pockets have emerged from higher bands which causes the increment of density of states that leads to decrease in the resistivity. The Fig.~\ref{appnd_fig2}-\ref{appnd_fig4} thus explains the double-peak feature seen in the recent experiments~\cite{anindya_thermo}}
	\label{appnd_fig4}
\end{figure}

	In the main text, we had shown that the presence of excitonic order leads to formation of 6 minibands out of the original tDBLG dispersion. Here, we present the detailed dispersion of these 6 bands and how they change with density around CNP in Fig.~\ref{appnd_fig2}-\ref{appnd_fig4}.
In Fig.~\ref{appnd_fig2}, we plot the dispersion of the 6 bands (3 conduction bands in top row and 3 valence bands in bottom row) at CNP as a color plot. The electron and hole Fermi surfaces are shown as solid lines. Note that the dispersion of the bands are not $C_3$ symmetric for individual $Q_i$'s. The e-h pocket that is coupled by the excitonic order is gapped out. Fig.~\ref{appnd_fig3} plots the same dispersion at $n/n_s=0.06$. One can see that the electron pockets have grown while the hole pockets have shrunk to zero. Finally, the Fig.~\ref{appnd_fig4} plots the dispersion at $n/n_s=0.10$. The hole pockets from the valence band have vanished and electron pockets are still present. But additional electron pockets appear in the conduction band. The extra density of states from these new pockets lead to a suppression of resistivity giving rise to the two-peak structure in the resitivity vs density data~\cite{anindya_thermo}. 

\end{appendices}

	% \bibliographystyle{apsrev} 
%	\bibliography{apssamp} 

\end{document}